\documentclass[aps,pra,twocolumn,superscriptaddress,a4paper]{revtex4}

\usepackage{amsmath} 
\usepackage{graphicx}
\def\degree{\mbox{$^\circ$}}
\def\ii{{\mathrm{i}}}
\def\ee{{\mathrm{e}}}

\def\sub#1{_{\scriptsize\mbox{#1}}}
\def\sur#1{^{\scriptsize\mbox{#1}}}
\def\vct#1{{\mathchoice{\mbox{\boldmath$#1$}}{\mbox{\boldmath$#1$}}%
  {\mbox{\scriptsize\boldmath$#1$}}{\mbox{\scriptsize\boldmath$#1$}}}}

\begin{document}
\title{Helical mode conversion using conical reflector}
\author{H. Kobayashi}
\affiliation{Department of Electronic and Photonic System Engineering,
Kochi University of Technology, Tosayamada-cho, Kochi 782-8502, Japan}
\author{K. Nonaka}
\affiliation{Department of Electronic and Photonic System Engineering,
Kochi University of Technology, Tosayamada-cho, Kochi 782-8502, Japan}
\author{M. Kitano}
\affiliation{Department of Electronic Science and Engineering, Kyoto
University, Kyoto 615-8510, Japan}

\date{\today}

\begin{abstract}
In a recent paper, Mansuripur \textit{et al.} 
[Phys. Rev. A {\bf 84}, 033813 (2011)] indicated and numerically verified
 the generation of the helical wavefront of optical beams using a
 conical-shape reflector. 
Because the optical reflection is largely free from chromatic aberrations, 
 the conical reflector has an advantage of being able to manipulate 
the helical wavefront with broadband light such as white light or short
 light pulses. 
In this study, we introduce geometrical understanding of the function of the
 conical reflector using the spatially-dependent geometric phase, or more specifically, the
 spin redirection phase. 
We also present a theoretical analysis based on three-dimensional matrix
 calculus and elucidate relationships of the spin, orbital, and total
 angular momenta between input and output beams. 
These analyses are very useful when designing other optical devices 
that utilize spatially-dependent spin redirection phases. 
Moreover, we experimentally demonstrate the generation of helical beams
 from an ordinary Gaussian beam using a metallic conical-shape
 reflector.
\end{abstract}
\maketitle
\section{Introduction}
When a physical system evolves along a path in parameter space
and returns to the initial state, its wavefunction acquires an 
additional phase factor that depends solely upon the path traced in parameter space. 
This phase factor, called the geometric phase, was first set out by
Berry~\cite{berry84:_quant_phase_factor_accom_adiab_chang}. 
There have been many manifestations of geometric phase in a
variety of physical systems~\cite{shapere89:_geomet_phases_in_physic}. 
In optics, there are two primarily types of geometric phase: 
one is the spin-redirection phase, which is induced by cyclic changes in
the propagation direction of a light beam~\cite{PhysRevLett.57.933,berry87:_inter_anhol_of_coiled_light,PhysRevLett.57.937,PhysRevLett.58.523,PhysRevLett.60.1584,PhysRevLett.60.1214,PhysRevLett.69.590,Galvez:99}. 
The other is the Pancharatnam-Berry
phase, which is associated with cyclic changes of
polarization~\cite{pancharatnam56:_proc,berry87:_adiab_phase_and_panch_phase}. 

In the course of studies on geometric phases, 
it has been shown that if a light wave is subjected to a transversely 
inhomogeneous state change with uniform initial and final states, 
the associated spatially-dependent geometric phases induce wavefront
reshaping~\cite{bhandari97:_polar_of_light_and_topol_phases,bomzon02:_space_varian_panch_berry_phase}. 
This method of wavefront reshaping is fundamentally different from 
that in typical optical devices, which utilize the optical path
difference, e.g., standard lenses and curved mirrors. 
Whereas the conventional methods suffer from chromatic aberrations, 
approaches based on geometric phases have the advantage of being able to
realize achromatic optical devices, because
the phase shift depends only upon the path in state space, and not upon wavelength. 

\begin{figure}[t]
\begin{center}
\includegraphics[width=5cm]{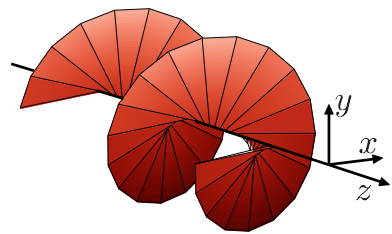}
\caption{Equiphase surface of helical beam with $l=+1$ propagating along $z$-axis.}
\label{fig:helical_phase_front_3D}
\end{center}
\end{figure}

One important application of spatially-dependent geometric
phase is the generation and mode conversion of light beams with
an optical vortex or helical beams. 
A beam is characterized by an integer $l$, called the helical
mode number. Its wavefront is composed of $|l|$ intertwined helical
wavefronts, with a handedness given by the sign of
$l$ (e.g., Fig.~\ref{fig:helical_phase_front_3D} shows the equiphase surface
of the helical beam with $l=+1$). 
It has been shown that each photon in the helical mode 
carries a quantized intrinsic orbital
angular momentum $l\hbar$, in addition to the spin-like angular
momentum $\pm\hbar$ associated with circularly polarized
waves~\cite{PhysRevA.45.8185}. 
Recently, helical beams have attracted growing interest, owing
to their possible use in optical trapping and manipulation of particles
and atoms~\cite{PhysRevLett.75.826}, multi-state information encoding
for optical communication, and quantum computation~\cite{mair01:_entan_of_orbit_angul_momen}.

In recent years, a specially manufactured half-wave plate called 
``$q$-plate'' has been used for the manipulation of helical beams~\cite{PhysRevLett.96.163905}. 
When a circularly polarized Gaussian beam is sent through a $q$-plate, 
the polarization state evolves along a path on the Poincar\'{e}
sphere that depends upon the azimuthal angle of the transversal plane 
and eventually becomes uniform but opposite circular polarization, 
as a result of half-wave retardation. 
Because of the associated Pancharatnam-Berry phase with the above-described state
evolution, the output light beam no
longer remains Gaussian, but becomes, instead, a helical beam. 
Unfortunately, the $q$-plate will only operate in the above-described
way at given wavelength, because the birefringent retardation must
correspond to exactly one half of the wavelength. 

More recently, Mansuripur and his colleagues proposed a new method of helical mode
conversion utilizing the spatially-dependent spin redirection
phase~\cite{PhysRevA.84.033813}. 
Using Maxwell-equation-based simulation, they showed the helical
mode generation from an ordinary Gaussian beam can be performed using a
conical-shape reflector due to spin-to-orbital angular momentum conversion. 
They also introduced a simplified
analysis based on the Jones calculus to elucidate the physics underlying
the reflection properties~\cite{mansuripur11:_spin_to_orbit_angul_momen}. 
Because the optical reflection is largely free from chromatic aberrations, 
 the conical reflector has the advantage of being able to manipulate 
the helical wavefront with broadband light such as white light or short
 light pulses. 

In this paper, we take Mansuripur's analysis of the conical reflector one
step further. 
We consider that the general helical beam, not the ordinary
Gaussian beam, is incident on the conical
reflector, and elucidate the relationship of spin, orbital,
and total angular momenta between input and output beams based on 
the three-dimensional matrix calculus. 
We also explicitly introduce the geometrical analysis using the spatially-dependent spin
redirection phases, which is implied in Ref.~\cite{PhysRevA.84.033813}. 
These analyses are very useful when designing other optical devices 
that utilize spatially-dependent spin redirection phases. 
Moreover, we design a metallic conical-shape reflector, 
and demonstrate helical mode generation
from an ordinary Gaussian beam. 
To the best of our knowledge, this is the first experimental
demonstration of the spin-to-orbital angular momentum conversion 
using the conical reflector. 

The remainder of this paper is organized as follows. 
In Sec.~\ref{sec:helic-mode-conv}, we present a theoretical analysis 
of helical mode conversion using a conical reflector 
based on the well-known matrix formula. Then, we study the principle of
operation from the viewpoint of the geometric phase or the
spin redirection phase. 
In Sec.~\ref{sec:experiment}, we describe our experimental setup and
results regarding helical mode conversion with the conical reflector. 
Conclusions are presented in Sec.~\ref{sec:conclusion}. 

\section{Theoretical analysis of helical mode conversion using
 conical reflector}
\label{sec:helic-mode-conv}
In this section, we describe the characteristics of helical
beams and present a theoretical introduction to helical mode
conversion using a conical reflector. 
Initially, we calculate its function using the
well-known matrix formula. 
Then, we show a geometrical interpretation which utilizes 
the spin redirection geometric phase. 

In what follows, we assume that all the reflectors have infinite
conductivity, which results in equal phase shifts 
for the S and P components of the polarization, and that the
propagation and reflection losses are negligible. 

\subsection{Helical beam and optical angular momentum}
In the paraxial approximation, 
a monochromatic helical wave with circular polarization propagated along
the $z$-axis is given by the wave vector $\vct{k}=k(0,0,\kappa)\sur{T}$, 
and the electric field vector $\vct{E}$:
\begin{align}
\vct{E}(\vct{x},t)&=
\tilde{E}_l(\vct{r})\ee^{\ii(\vct{k}\cdot\vct{x}-\omega t)}\vct{e}_\pm+\text{c.c.} \nonumber\\
&=E_0(r)\ee^{\ii\kappa(l\phi+kz)}\ee^{-\ii\omega t}\vct{e}_\pm+\text{c.c.}
,  \label{eq:1}
\end{align}
where $k$ is the wavenumber, $\kappa$ represents the sign of 
the propagation direction ($\kappa=\pm1$ corresponds to the $\pm z$ direction), 
$\vct{x}\equiv(x,y,z)$, $\vct{r}\equiv(x,y)$, 
$r$ and $\phi$ are the polar coordinates in the $x$-$y$ plane, 
the integer $l$ represents the helical mode number, and the unit complex vectors 
$\vct{e}_\pm=(1,\pm\ii,0)\sur{T}/\sqrt{2}$ represent the circular
polarizations.  
The helicity $\sigma$ is calculated using Eq.~(\ref{eq:1}) as $\pm\kappa$ 
(see Appendix \ref{sec:helic-vect-helic}). 

\begin{figure}[tbp]
\begin{center}
\includegraphics[width=8cm]{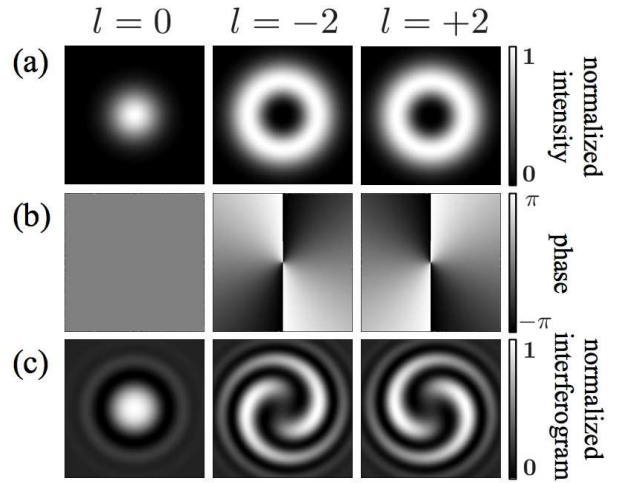}
\caption{Transverse profile of fundamental Gaussian beam ($l=0$) and 
Laguerre Gaussian beam with $l=\pm 2$. 
(a) Intensity profile. (b) Phase
 profile. (c) Interferogram with a spherical wave.}
\label{fig:helical_phase_front}
\end{center}
\end{figure}

With certain choices of radial profile, Eq.~(\ref{eq:1}) corresponds to
the well-known Laguerre-Gaussian modes. 
The intensity and phase profiles of this mode are shown in
Fig.~\ref{fig:helical_phase_front}(a) and (b), respectively. 
The phase profile, as shown in Fig.~\ref{fig:helical_phase_front}(b), can
not be measured directly, however, we can observe the phase profile via the interferogram between
the helical beam and a spherical wave [see
Fig.~\ref{fig:helical_phase_front}(c)]. 
For non-helical waves ($l=0$), the resulting interferogram consists of
concentric circular fringes. 
For helical waves, however, the interferogram takes the form of spirals
(double spiral in the case of $|l|=2$), 
with a handedness that depends upon the sign of the helical mode: 
clockwise (counterclockwise) outgoing spirals correspond to a positive
(negative) mode number. 

In addition to the spin angular momentum $\sigma\hbar$ along the propagation direction
per photon, 
the light beam of Eq.~(\ref{eq:1}) carries an orbital angular momentum
$l\hbar$ per photon~\cite{PhysRevA.45.8185}. 
The total angular momentum along the $z$-axis,
$J_z$, can be represented by
\begin{align}
J_z&=L_z+S_z\nonumber\\
&=\hbar\kappa(l+\sigma),  \label{eq:2}
\end{align}
where $L_z$ and $S_z$ are the $z$ component
of the orbital and spin angular momentum, respectively.

\subsection{Reflection matrix of conical reflector and helical mode conversion}

Let us consider the reflection of a monochromatic plane wave with wave vector
$\vct{k}$ from a perfect mirror M, which can be
characterized by a unit vector $\vct{m}$ normal to its surface. 
The function of the reflection, 
identified by a matrix ${\mathcal M}$, can be represented by the 
rotation of the angle $\pi$ around vector $\vct{m}$ followed 
by inversion (parity transformation) as follows:
\begin{align}
{\mathcal M}
=-{\mathcal R}(\pi,\vct{m})
={\mathcal I}-2\vct{m}\vct{m}\sur{T},  \label{eq:3}
\end{align}
where ${\mathcal R}(\theta,\vct{a})$ is the matrix for the rotation of
$\theta$ in the right-handed sense about an axis $\vct{a}$, 
${\mathcal I}$ is the unit matrix, and $\vct{m}\sur{T}$ is the
transpose of vector $\vct{m}$~\cite{PhysRevLett.60.1584,PhysRevLett.69.590}. 
More generally, because the rotation matrix ${\mathcal R}$ is a member of SO(3), 
a sequence of $N$ reflections with individual reflection matrices
${\mathcal M}_i$ $(i=1,\cdots,N)$ can be represented by a single
rotation matrix as follows:
\begin{align}
{\mathcal M}=\prod_{i=1}^N{\mathcal M}_i
=(-1)^N{\mathcal R}(\theta\sub{T},\vct{a}\sub{T}),  \label{eq:4}
\end{align} 
where $\theta\sub{T}$ is the equivalent angle of rotation around an axis $\vct{a}\sub{T}$. 

\begin{figure}[tbp]
\begin{center}
\includegraphics[width=8.5cm]{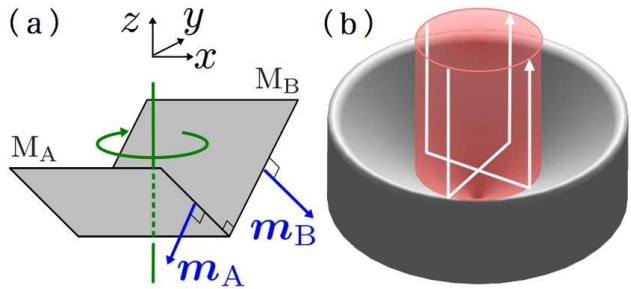}
\caption{
(a) Two-dimensional (dihedral) corner reflector. Two perfect
 mirrors, M$\sub{A}$ and M$\sub{B}$, 
 constituting the corner reflector are represented by the 
 vectors normal to their surfaces, $\vct{m}\sub{A}$ and
 $\vct{m}\sub{B}$, respectively. 
(b) Conical reflector. This reflector is made as the solid of
 revolution generated by the rotation of the dihedral corner reflector
 around $z$-axis.}
\label{fig:two_conical_mirrors}
\end{center}
\end{figure}

First, we calculate the reflection matrix of 
a two dimensional (dihedral) corner reflector 
that consists of two mutually perpendicular mirrors, 
M$\sub{A}$ and M$\sub{B}$ with normal vectors $\vct{m}\sub{A}=(0,-1,-1)\sur{T}/\sqrt{2}$ and
$\vct{m}\sub{B}=(0,1,-1)\sur{T}/\sqrt{2}$, respectively 
[see Fig.~\ref{fig:two_conical_mirrors}(a)]. 
The reflection matrix ${\mathcal M}_0$ for the dihedral corner reflector
is given by
\begin{align}
{\mathcal M}_0&={\mathcal M}\sub{B}{\mathcal M}\sub{A}
={\mathcal M}\sub{A}{\mathcal M}\sub{B}
=
\begin{pmatrix}
1 & 0 & 0 \\
0 & -1 & 0 \\
0 & 0 & -1
\end{pmatrix}
,  \label{eq:5}
\end{align}
which corresponds to the $\pi$ rotation around the $x$-axis. 

Next, we consider a conical reflector, which is a solid of revolution
generated by the rotation of a dihedral corner
reflector around the $z$-axis, as shown in
Fig.~\ref{fig:two_conical_mirrors}(b). 
The axial-symmetric parallel light beam enters along the $z$-axis and
is reflected on the reflector. 
If the beam diameter is large enough relative to the wavelength, the
incident beam can be considered as a bundle of rays. 
Each ray, which is parallel to the $z$-axis, can be labeled by
$\vct{r}=(x,y)$. 
The polarization vector can be assigned to the incident ray in the form of the
transverse electric field vector $\vct{\tilde{E}}\sub{in}(\vct{r})$. 
Similarly, the output polarization carried by
the ray can be represented by $\vct{\tilde{E}}\sub{out}(\vct{r})$. 
The diameter of each ray is assumed to be still much larger than the
wavelength, and the diffraction associated with propagation can be neglected. 
Thus, the effect of the conical reflector can be analyzed on a
ray-by-ray basis. 

For an incident ray at $\vct{r}=(x,y)$, the conical reflector can be
replaced by a tangential dihedral corner reflector that is obtained by
the rotation of $M_0$ about the $z$-axis by
$\phi=\tan^{-1}(y/x)$. 
Thus, the input ray is transformed according 
to the reflection matrix
\begin{align}
{\mathcal M}\sub{CR}(\phi)
=&{\mathcal R}(-\phi,\vct{e}_z)
{\mathcal M}_0
{\mathcal R}(\phi,\vct{e}_z)  \nonumber\\
=&
\begin{pmatrix}
\cos 2\phi & \sin 2\phi & 0 \\
\sin 2\phi & -\cos 2\phi & 0 \\
0 & 0 & -1
\end{pmatrix}
,  \label{eq:6}
\end{align}
which corresponds to the rotation of $\pi$ around the vector
$\vct{n}=(\cos\phi,\sin\phi,0)$. 
Moreover, the input ray at $\vct{r}$ is relocated to the transverse position $-\vct{r}$. 
Therefore, the relationship between the input ray and the output ray is given by
\begin{align}
\vct{\tilde{E}}\sub{out}(-\vct{r})
&=\mathcal{M}\sub{CR}(\phi)\vct{\tilde{E}}\sub{in}(\vct{r}).  \label{eq:7}
\end{align}

Next, we consider a helical wave with a mode number $l$ 
and a circular polarization ($\sigma=\pm 1$) propagating in the $-z$
direction that is incident on the conical reflector. 
The input field can be represented by
\begin{align}
%\vct{k}\sub{in}&=k(0,0,-1)\sur{T}, 
\vct{E}\sub{in}(\vct{x},t)
&=\vct{\tilde{E}}\sub{in}(\vct{r})\ee^{-\ii(kz+\omega t)}+\text{c.c.} \nonumber\\
&=E_0(r)\ee^{-\ii(l\phi+kz+\omega t)}\vct{e}_\mp+\text{c.c.},  \label{eq:8}
\end{align}
where the unit complex vectors $\vct{e}_-$ and
$\vct{e}_+$ correspond to the right ($\sigma=+1$) and the left
($\sigma=-1$) circular polarization, respectively. 

From Eqs.~(\ref{eq:7}) and (\ref{eq:8}), 
the output field from the conical reflector is
\begin{align}
%\vct{k}\sub{out}&=-\vct{k}\sub{in}=k(0,0,1)\sur{T},
\vct{E}\sub{out}(\vct{x},t)
&=\ee^{-\ii
 2\sigma\phi}\tilde{\vct{E}}\sub{in}(-\vct{r})\ee^{\ii(kz-\omega t)}+\text{c.c.}  \nonumber\\
&=E_0(r)\ee^{\ii[-(l+2\sigma)\phi+kz-\omega t]}\vct{e}_\pm+\text{c.c.}\nonumber\\
&=\tilde{E}_{-l-2\sigma}(\vct{x})
\ee^{-\ii\omega t}\vct{e}_\pm+\text{c.c.}  \label{eq:9}
\end{align}
From this equation, we see that the output wave is uniformly
 circularly polarized ($\sigma=\pm1$); however, its wavefront has acquired a
 nonuniform phase factor $-2\sigma\phi$ that depends upon the input helicity. 
Because the conical reflector inverts the wave vector,
$\vct{k}\sub{out}=-\vct{k}\sub{in}$, the output helicity is conserved 
and the input helical mode number $l$ is converted to
$-l-2\sigma$. 

\begin{figure}[htbp]
\begin{center}
\includegraphics[width=8cm]{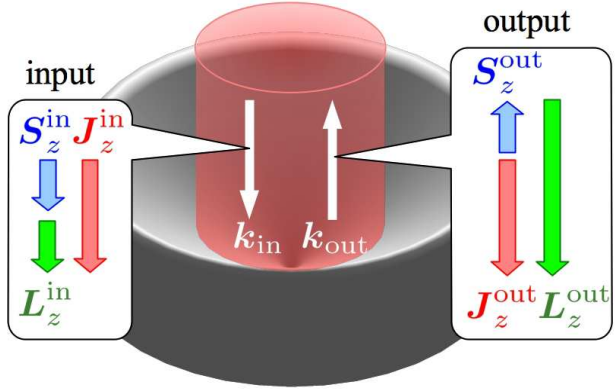}
\caption{The relationship between input and output angular momentum
 vectors. $\vct{S}_z=S_z\vct{e}_z$,
 $\vct{L}_z=L_z\vct{e}_z$, 
and $\vct{J}_z=J_z\vct{e}_z$ represent $z$
 components of the spin, orbital, and total angular momentum vectors,
 respectively.}
\label{fig:angular_momentum}
\end{center}
\end{figure}

From Eqs.~(\ref{eq:8}) and (\ref{eq:9}), the relationship between the
$z$ components of the input and output angular momentum can be calculated as 
\begin{align}
S_z\sur{out}&=-S_z\sur{in},  \label{eq:10}\\
L_z\sur{out}&=L_z\sur{in}+2S_z\sur{in},  \label{eq:11}\\
J_z\sur{out}&=J_z\sur{in},  \label{eq:12}
\end{align}
where the superscripts ``in'' and ``out'' denote the input and output
(Figure~\ref{fig:angular_momentum} illustrates the above relationships). 
Equation~(\ref{eq:12}) shows that the $z$ component of 
total angular momentum is conserved
under reflection and that no angular momentum is transferred from
the incident light field to the conical reflector. 
Thus, the conical reflector acts only as a ``catalyst'' for the
interconversion between spin and orbital angular momenta. 
This result agrees well with those of previous studies, which indicate that an axially
symmetric perfect electrical conductor cannot acquire any
angular momentum along its axis of symmetry
~\cite{konz03:_geomet_absor_of_elect_angul_momen,nieminen04:_commen_geomet_absor_of_elect_angul_momen,PhysRevA.84.033813}. 

\subsection{Geometrical understanding of helical mode conversion using conical reflector}
The spatially-dependent phase factor discussed in the previous section 
is not induced by the optical path length difference. 
In fact, if the light wave is incident along the $z$-axis, all optical
paths under reflection from the conical reflector are obviously the
same. Instead, we have utilized the so-called spin redirection phase, which is one of
the manifestation of the geometric phase. 
Its magnitude depends only upon the trajectory of
the state vector traced in state space. 
In the case of optical systems in which the light propagates along a
three-dimensional path, 
the state vector is the helicity vector with unit length 
and the state space is the spherical
surface pointed by the helicity vector (further information regarding the helicity and the
helicity vector is presented in Appendix~\ref{sec:helic-vect-helic}). 

When the helicity vector is gradually changed as in the case of
propagation through a helically-wound optical
fiber~\cite{PhysRevLett.57.933,PhysRevLett.57.937}, 
the helicity is conserved and the temporal change of
the helicity vector can be represented by
\begin{align}
\vct{\sigma}(t)=\frac{\sigma\vct{k}(t)}{k}.  \label{eq:13}
\end{align}
The geometric phase is proportional to the surface area of the sphere 
enclosed by the
closed path $\vct{\sigma}(t)$, where $\vct{\sigma}(T)=\vct{\sigma}(0)$. 

In an optical system in which the light is reflected by perfect mirrors, 
however, the helicity is inverted after each reflection [see Eq.~(\ref{eq:4})]. 
The helicity vector after $n$ reflections can be represented as follows:
\begin{align}
\vct{\sigma}_n&=\frac{\sigma_n\vct{k}_n}{k}
=\frac{(-1)^n\sigma_0\vct{k}_n}{k},  \label{eq:14}
\end{align}
where 
$\vct{\sigma}_n$ and $\vct{k}_n$ are the helicity vector and the wave vector
after the $n$-th reflection.
Thus, we consider modified $\vct{k}$ vectors as a state vector
defined by 
\begin{align}
\tilde{\vct{k}}_n=\frac{(-1)^n\vct{k}_n}{k}.  \label{eq:15}
\end{align}
If we measure the helicity of the photon with respect to
$\tilde{\vct{k}}$ vectors, 
it is conserved under perfect reflections, as in the case of gradual
change~\cite{PhysRevLett.58.523,PhysRevLett.69.590,Galvez:99}. 

The spin redirection phase $\gamma$ is proportional to the area $\Omega$
enclosed by the path formed by discrete points on the $\tilde{\vct{k}}$-sphere 
connected by geodesics: 
\begin{align}
\gamma=-\sigma\Omega.  \label{eq:16}
\end{align}
The sign of this phase will depend upon the helicity of the light spin. 

\begin{figure}[t]
\begin{center}
\hspace*{-0.4cm}
\includegraphics[width=9cm]{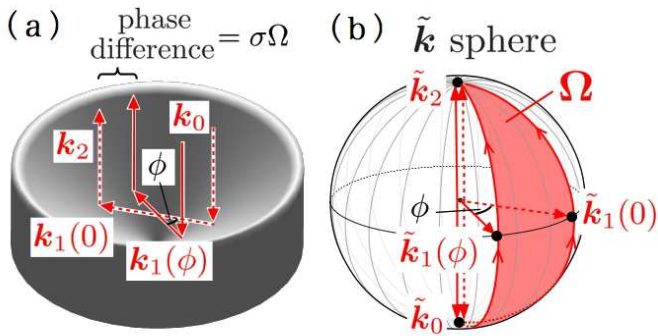}
\caption{Reflection from the surface of conical reflector. (a) The
 changing of the wave vector during the reflection on the conical
 reflector. 
(b) Change of the modified wave vector $\tilde{\vct{k}}_n$ on
 $\tilde{\vct{k}}$-sphere.}
\label{fig:cone_mirror}
\end{center}
\end{figure}

Figure~\ref{fig:cone_mirror} shows the wave vectors $\vct{k}_n$ and the modified wave
vectors $\vct{\tilde{k}}_n$ of a beam that undergoes two reflections on the conical
reflector. 
The input beam propagating along the $z$-axis 
is subjected to two perfect reflections, and finally, 
its direction is inverted; $\vct{k}_0=-\vct{k}_2$. 
The wave vector $\vct{k}_1(\phi)$ 
in the middle of the two reflections, however, depends upon 
the azimuthal angle $\phi$ in the $x$-$y$ plane, i.e., 
the path of state evolution is spatially dependent. 
Thus, the associated spin redirection phases induce wavefront reshaping.
The area $\Omega$ is proportional to the azimuthal angle $\phi$ and we obtain the
spin redirection phase $\gamma=-2\sigma\phi$ [see
Fig.~\ref{fig:cone_mirror}(b)], 
which is equal to the phase factor calculated in the previous section [Eq.~(\ref{eq:9})]. 
This geometrical understanding is very clear, and is useful when
designing optical devices that utilize spatially-dependent geometric phase. 

\section{Experiment}
\label{sec:experiment}
\begin{figure}[t]
\begin{center}
\hspace*{1.5cm}
\includegraphics[width=6cm]{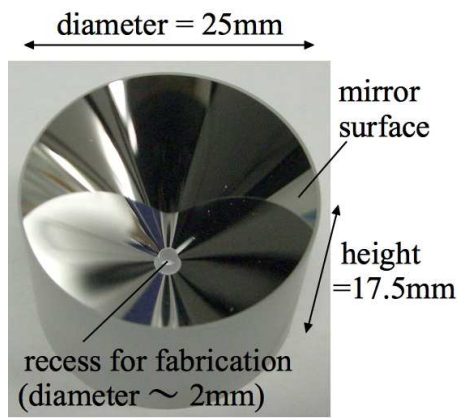}
\caption{Fabricated conical reflector. For the grinding 
process of the fabrication, there remains a tiny recess at the
 center of the conical reflector.}
\label{fig:cone_mirror_picture}
\end{center}
\end{figure}

\begin{figure}[t]
\begin{center}
\hspace*{-0.4cm}
\includegraphics[width=9cm]{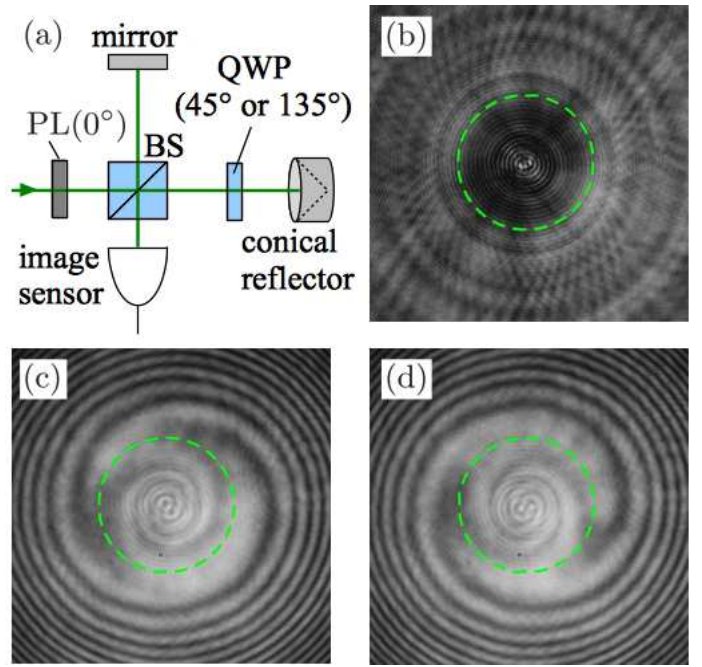}
\caption{Experimental setup and results. (a) Michelson
 interferometer to observe the spiral interferogram of the helical
 mode. (b) Intensity distribution of the reflected wave from the
 conical reflector. (c), (d) Interferograms generated by the reflected beam
 from the conical reflector and the quasi-spherical wave. 
 The area encircled by the green dashed line corresponds to the beam
 diffracted from the recess at the bottom of the conical reflector.
 There are some deformations of wavefronts. We could remove this artifact
 by using a perfect conical reflector without a recess.}
\label{fig:laguerre_gaussian_results}
\end{center}
\end{figure}

To demonstrate the helical mode conversion using the conical reflector, 
we designed an aluminum-coated conical reflector that is reflective at visible wavelengths. 
Figure~\ref{fig:cone_mirror_picture} shows the fabricated conical
reflector (Natsume Optical Corporation), 
which is 25 mm in diameter and 17.5 mm in height. 
The surface accuracy of the conical surface 
is less than $5\lambda$. 
Unfortunately, the grinding process for the conical surface results in a
small recess in the center of the reflector.

To verify the spiral wavefront shape of the light emerging from
the conical reflector, we constructed a Michelson interferometer, as shown in
Fig.~\ref{fig:laguerre_gaussian_results}(a). 
After beam shaping using a single mode optical fiber, a 532-nm laser
beam with a beam-waist radius of approximately 5 mm is sent through the horizontal
linear polarizer (PL). 
Next, the beam is split into two beams by a non-polarizing beam splitter
(BS). 
One beam is sent to the conical reflector, which we call the signal
beam, 
and the other is used as a reference. 
The initial horizontal polarization of the signal beam 
is changed to right or left circular
polarization using a quarter-wave plate with a $45\degree$ or
$135\degree$ angle of the fast axis, 
and is then incident on the conical reflector. 
The beam reflected from the conical reflector is then sent through the
quarter-wave plate again and its polarization is reverted to the original
horizontal polarization. 
Finally, the signal beam was superimposed with the reference beam by the
BS to yield the interferogram, which was observed with an image sensor. 

First, we observed the intensity profile of the reflected beam from the
conical reflector by blocking the reference beam
[Fig.~\ref{fig:laguerre_gaussian_results}(b)]. 
The intensity profile was found to have the
doughnut-like shape expected for a helical mode [see
Fig.~\ref{fig:helical_phase_front}(a)]. 
Within the region encircled by the green dashed line in
Fig.~\ref{fig:laguerre_gaussian_results}(b), however, 
there are spurious waves diffracted from the center recess of the reflector. 

Next, we observed the interferogram between the signal and reference
beams. 
Because the optical length of the reference arm is much larger than that
of the signal arm, 
the wavefront of the reference beam can be considered approximately
spherical. 
Figures~\ref{fig:laguerre_gaussian_results}(c) and (d) show the acquired
images of the interference patterns for the quarter waveplate 
with $45\degree$ [Fig.~\ref{fig:laguerre_gaussian_results}(c)] and
$135\degree$ [Fig.~\ref{fig:laguerre_gaussian_results}(d)] angle of the fast
axis. 
The inner region of the green dashed lines includes the light affected
by the diffraction from the center recess. 
The double spirals outside the dashed line in
Fig.~\ref{fig:laguerre_gaussian_results}(c) and (d) 
show unambiguously that the wavefront
of the light reflected from the conical reflector has helical wavefronts 
with mode number $l=\pm 2$, and that the input polarization on the
conical reflector can control the sign of the mode conversion. 

\section{Conclusion}
\label{sec:conclusion}
In this paper, we have elucidated the relationship of spin, orbital, and
total angular momenta between input and output beams on the conical
reflector based on the three-dimensional matrix calculus. 
We have also  presented a theoretical analysis of the function of a
conical reflector using the spatially-dependent spin redirection phase. 
Moreover, we have experimentally verified helical mode conversion using a fabricated conical reflector. 
The experimental results show that the conical reflector induces 
$2$-raising or $2$-lowering helical mode conversion, which depends upon the
input helicity. 

Because the actual reflectors, which have finite conductivities, 
induce unequal phase shifts for the S and P components
of the polarization, the conservation of helicity is partly violated. 
However, it is possible to compensate for the imperfection 
by placing a circular polarizer in front of the conical reflector. 
In this case, the output light field will have the expected 
helical wavefront, regardless of the wavelength, but at the price of some optical loss. 
Thus, unlike previous methods, the conical reflector has the advantage of
allowing the manipulation of the helical wavefront of light 
with a wide spectral range, e.g.,
white light and short light pulses. 
Our demonstration serves as a foundation for the design of new
achromatic optical devices that utilize spatially-dependent spin
redirection phase. 

\begin{acknowledgements}
We thank Shuhei Tamate in Kyoto University, Japan for providing useful comments
 and suggestions.
\end{acknowledgements}

\begin{appendix}
\section{Helicity vector and helicity of polarized plain wave}
\label{sec:helic-vect-helic}
Let us consider a monochromatic wave with an electric field
\begin{align}
\vct{E}(\vct{x},t)=\vct{\tilde{E}}(\vct{x})\ee^{-\ii\omega t}+\text{c.c.},  \label{eq:17}
\end{align}
where $\vct{\tilde{E}}(\vct{x})$ is the complex electric field and 
$\omega$ is the angular frequency. 
The helicity vector of the above wave field can be defined as
\begin{align}
\vct{\sigma}&\equiv
\ii\frac{\vct{\tilde{E}}(\vct{x})\times\vct{\tilde{E}}^*(\vct{x})}
{|\vct{\tilde{E}}(\vct{x})|^2}.  \label{eq:18}
\end{align}
The magnitude $|\vct{\sigma}|$ corresponds to the
polarization ellipticity and the direction indicates the handedness of 
the temporal rotation of $\vct{E}$. 

For an arbitrary polarized plane wave propagating
along the $z$-axis, 
\begin{align}
\vct{\tilde{E}}(\vct{x})&=E_0\ee^{\ii\vct{k}\cdot\vct{x}}
\begin{pmatrix}
\ee^{\ii\phi_1}\cos\theta\\
\ee^{\ii\phi_2}\sin\theta\\
0
\end{pmatrix}
,  \label{eq:19}
\end{align}
with $0\leq\theta\leq\pi/2$, $0\leq\phi_1\leq 2\pi$, and
$0\leq\phi_2\leq 2\pi$, the helicity vector is calculated as
\begin{align}
\vct{\sigma}&=\sin(\phi_2-\phi_1)\sin(2\theta)\vct{e}_z,  \label{eq:20}
\end{align}
where $\vct{e}_z$ is the unit vector along the $z$-axis. 
The magnitude of Eq.~(\ref{eq:20}) is equal to the Stokes parameter $S_3$,
which corresponds to the polarization ellipticity. 
The rotation direction of the electric field vector follows the 
so-called right-hand screw rule around $\vct{\sigma}$. 

Moreover, even if an arbitrary three-dimensional rotation ${\mathcal R}$
is applied to $\vct{\tilde{E}}$, $\vct{\sigma}$ is conserved
in the sense that the rotation in $\vct{\tilde{E}}$ induces the same rotation to $\vct{\sigma}$:
\begin{align}
\vct{\sigma}^\prime&=
\ii\frac{{\mathcal R}\vct{\tilde{E}}(\vct{x})\times{\mathcal R}\vct{\tilde{E}}^*(\vct{x})}
{|{\mathcal R}\vct{\tilde{E}}(\vct{x})|^2}  \nonumber\\
&=\ii{\mathcal R}\frac{\vct{\tilde{E}}(\vct{x})\times\vct{\tilde{E}}^*(\vct{x})}{|\vct{\tilde{E}}(\vct{x})|^2}
={\mathcal R}\vct{\sigma}.  \label{eq:21}
\end{align}
Thus, the helicity vector of the arbitrary electric field has the
property described in the first paragraph of this appendix. 

Because of the transversality condition, $\vct{k}\cdot\vct{E}=0$, 
$\vct{\sigma}$ is parallel or anti-parallel to $\vct{k}$ 
(we can confirm this property from $\vct{k}\times\vct{\sigma}=0$). 
When $\vct{\sigma}$ is parallel (anti-parallel) to $\vct{k}$, 
we call it right-handed (left-handed) circularly polarized light. 
To determine the handedness of the polarization, we define the
helicity $\sigma$ by
\begin{align}
\sigma&\equiv\frac{\vct{k}\cdot\vct{\sigma}}{k},  \label{eq:22}
\end{align}
where $k$ is the wavenumber of the plane wave. 
The magnitude and the sign of $\sigma$ indicate the ellipticity and the
handedness of the polarization, respectively. 
From Eq.~(\ref{eq:22}), the helicity vector can be represented using
the helicity and the wave vector as
\begin{align}
\vct{\sigma}=\frac{\sigma\vct{k}}{k}.  \label{eq:23}
\end{align}
\end{appendix}

\end{document}